Little-Parks Effect Governed by Magnetic Nanostructures with Out-of-Plane Magnetization

M. C. de Ory[1], V. Rollano[1], A. Gomez[2], M. Menghini[1], A. Muñoz-Noval[3], E. M. Gonzalez[1,3] and J. L. Vicent[1,3,]
[1] IMDEA-Nanociencia, Cantoblanco, E-28049 Madrid, Spain
[2] Centro de Astrobiología (CSIC-INTA), Torrejón de Ardoz, E-28850 Madrid, Spain
[3] Departamento Física de Materiales, Universidad Complutense, E-28040 Madrid, Spain

**Abstract**
Little-Parks effect names the oscillations in the superconducting critical temperature as a function of the magnetic field. This effect is related to the geometry of the sample. In this work, we show that this effect can be enhanced and manipulated by the inclusion of magnetic nanostructures with perpendicular magnetization. These magnetic nanodots generate stray fields with enough strength to produce superconducting vortex-antivortex pairs. So that, the L-P effect deviation from the usual geometrical constrictions is due to the interplay between local magnetic stray fields and superconducting vortices. Moreover, we compare our results with a low-stray field sample (i.e. with the dots in magnetic vortex state) showing how the enhancement of the L-P effect can be explained by an increment of the effective size of the nanodots.

**Introduction**

After the finding of the quantization of magnetic flux trapped in a superconducting cylinder (1, 2), W. A. Little and R. D. Parks (3) realized that the superconducting critical temperature must be a periodic function of the magnetic flux. The Little-Parks effect (L-P effect in the following) is a fundamental superconducting effect related to the interplay between the Cooper pairs density and the free energy of the superconducting state (4, 5) at the transition temperature. Later on, Little and Parks explored what happens to the flux quantization in a multiply-connected superconductor (6). They showed that a maximum in critical temperature ($T_c$) yields a minimum in the resistance (with $R(H)$ measured at constant temperature), so L-P effect produces magnetoresistance oscillations. In these multiply-connected superconductors the dimensions of the superconducting areas (wires) and holes are crucial.

This subtle and basic superconducting effect has been used as the footprint of superconductivity in mesoscopic systems (7 – 9) and in very complex scenarios as, for example, detecting spatial modulation of superconductivity in 2D strongly correlated electron materials with coexistence of incommensurate and commensurate charge density waves (10) or in disorder-induced superconductor-insulator transitions (11, 12).There are some experimental regimes which can compete or mask the L-P effect; weak-link nanostructures are a clear example, since they are able to generate magnetoresistance oscillations as a function of the applied magnetic field in a broad range of temperatures (13). To circumvent this issue, the system has to be in the superconducting wire network regime; i. e. the condition $W < 1.84\,\xi(T)$ has to be fulfilled 14, 15), being $W$ the width of the superconducting wire and $\xi(T)$ the superconducting coherence length. Consequently, L-P effect can be found at temperatures close to $T_c$.

The interplay between superconducting vortices and L-P effect has to be considered too.

So far, several studies have addressed the role of the vortex movement in systems which also present the L-P effect; for instance, Berdiyorov et al. (16) show that the vortex effect should be considered in parallel with the Little-Parks effect; Sochnikov et al. (17) study the interaction between thermally excited moving vortices and the oscillating persistent currents induced in the superconducting loops; Welp et al. (18) and Gomez et al. (19) study the interplay between vortex pinning and L-P effect.

One interesting scenario is L-P effect in hybrid superconductors shaped by periodically arranged magnetic nanostructures. In this type of systems, the magnetic islands mimic the "holes" of the multiply-connected superconductors. Since superconductivity is depressed in the vicinity of the magnetic islands , the network consists of superconducting channels (wires) in between those areas where the superconductivity is strongly depleted by the array of magnetic dots. Quite a few works have dealt with this type of samples (array of magnetic dots on top or embedded in superconducting films), see review (20) and references therein; but to our knowledge, only few of them have focused on the experimental study of L-P effect in these nanostructures in the wire network regime (19, 21-23). The aim of this work is to explore whether or not the magnetism of the dots plays any role in the L-P effect. Furthermore, we will show that the stray field profile generated by the magnetic dots deeply modifies the L-P effect, depending upon the presence of vortices and antivortices pairs.

## Experimental

Two different hybrid samples consisting of arrays of magnetic nanodots covered by a superconducting film of superconducting niobium have been fabricated. Both systems have the same unit cell and are in direct contact with the superconducting film. The only difference between them is the different magnetic material used in the nanodots. One of them, SN sample in the following, contains an array of Ni dots in vortex state while in the second one, SCP sample in the following, the magnetic material consists of a Co/Pd multilayers, that ensures out of plane magnetization.

Both arrays were patterned using standard Electron Beam Lithography (PMMA resist) on a silicon (100) substrate. The nanodots diameter is 200 nm and they are arranged in a square array with side 400 nm. After the lithography, in the case of the SCP sample, a 40 nm thick [0.4 nm Co/ 0.6 nm Pd]$_{40}$ multilayer with a 3 nm Pd capping was sputtered on top, followed by a lift off process. The Co/Pd multilayers were obtained by magnetron sputtering using two targets and rotatable sample holder with Ar pressure of 12 mTorr (base pressure was $5 \times 10^{-8}$ Torr). In the case of the SN sample, a single Ni layer 40 nm thick was deposited in similar sputtering conditions. Finally, a 100 nm thick Nb film was deposited on top of both arrays using magnetron sputtering. For electric transport measurements, an 8-terminal cross-shaped bridge was defined on the Nb film using Optical Lithography and Ar/SF$_6$ (1:2) Reactive Ion Etching. The transport measurements were carried out using a commercial He cryostat housing a superconducting solenoid and a variable temperature insert. The temperature was controlled by a commercial temperature controller. The temperature stability is better than 1 mK. The experimental transport data are taken by the usual four probe dc technique.

Magnetic configurations at remanence were obtained from micromagnetic simulations performed with the finite difference code MuMax3 (24). These micromagnetic simulations were performed using 800×800×40 nm$^3$ unit cell that contains four dots, and with periodic boundary conditions to generate the complete array. The unit cell was discretized into 3×3×1 nm$^3$. For the Ni dots sample typical material parameters have been used: M$_S$ = 4.8 × 10$^5$ Am$^{-1}$, A = 6 × 10$^{-12}$ Jm$^{-1}$ and K = 0 Jm$^{-3}$ being $M_S$ the saturation magnetization, A the exchange constant, and K the in-plane anisotropy. Ni magnetocrystalline anisotropy was neglected due to its relative strength to the exchange stiffness (25). CoPd dots simulations were performed with material parameters of $M_S = 8.8 \times 10^5$ Am$^{-1}$, A = $3 \times 10^{-11}$ Jm$^{-1}$ and K = $2.40 \times 10^5$ Jm$^{-3}$ (26, 27). Stray field calculations were carried out numerically at 50 nm height from the top surface of the dots.

## Results and Discussions

Magnetoresistance measurements allow to characterize the superconducting vortex dynamics in hybrid systems made of superconducting films on top of magnetic arrays, see the review (20) and references therein. Superconducting vortex dynamics shows sharp and well define minima when vortex lattice is commensurate with the array, i.e., when the density of the vortex lattice is an integer number of the density of magnetic dots. Besides, these magnetoresistance minima permit to distinguish among different remanent magnetic states (28).

In the present work, two hybrid systems have been studied: Sample SN (Nb thin film on top of array of Ni nanodots) and Sample SCP (Nb thin film on top of array of Co/Pd multilayered dots).

Figure 1 (a) and (b) show the magnetoresistance minima measured in SN and SCP samples respectively. In sample SN, we observe that the magnetoresistance curve is symmetrical with respect to the direction of the magnetic field as usual (20). Otherwise, in sample SCP the magnetoresistance curve is shifted along the magnetic field axis resulting in an asymmetric profile. This shifting is well understood (29) and it is the footprint of the formation of vortex – antivortex pairs due to the rise of the stray magnetic fields emerging from the nanodots. In brief, Co/Pd multilayers show a rich magnetic behavior (28) which depends on the Co and Pd layers thickness. In our case (sample SCP), the Co and Pd layers thicknesses are chosen to generate a strong out-of-plane anisotropy (30). Therefore, when the remanent magnetization pointing out in the upward direction, the magnetic stray field in between dots point downwards, and superconductivity diminishes. If this field is high enough (31), it can generate vortex antivortex pairs, placing antivortices in the space between the dots. As a consequence, superconducting vortices are present regardless the presence of an external field (31), giving account for the asymmetric profile in the magnetoresistance. This effect can be compensated by applying an external magnetic field in the same direction of the remanent magnetization, hence retrieving superconductivity in the interstitial sites. The same analysis can be done when the magnetization in the remanent magnetic state points downwards.

In conclusion, the symmetric magnetoresistance curve obtained in the SN magnetic/superconducting system indicates that the magnetic state of the array corresponds to the magnetic vortex state at remanence (32) with negligible magnetic stray field. Conversely, the hybrid system SCP shows a strong out-of-plane magnetization; hence, the magnetic stray field deeply modifies the superconducting behavior.

The superconducting coherence length ($\xi(T)$) is a crucial parameter in our study that can be obtained by magnetotransport characterization. $\xi(0)$ is extracted from the slope of $H_{c2}$ (upper critical field) vs. temperature ($T$) at the critical temperature ($T_{c0}$) following the well-known WHH approach (33, 34):

$$S \equiv -\mu_0 \left(\frac{dH_{c2}}{dT}\right)_{T=T_{c0}} \qquad (1)$$

$$\xi(0) = 1.81 \times 10^{-8} \ [T_{c0} \ S]^{-1/2} \qquad (2)$$

The obtained $\xi(0)$ values using equations (1) and (2) are $\xi(0) = 9.2$ nm for the SN sample and $\xi(0) = 8.6$ nm for the SCP sample.

Once $\xi(0)$ is obtained, the temperature dependence of the coherence length is easily calculated using:

$$\xi(T) = \frac{\xi(0)}{\sqrt{1 - T/T_{c0}}} \qquad (3)$$

In our case, the key measurement to detect the L-P effect is the experimental determination of the (H, T) phase diagram of a superconducting network (35). First of all, we have to establish the temperature interval in which the wire network condition is fulfilled; i.e., W<1.84$\xi$(T). In principle, W is a geometrical parameter given by the distance between the magnetic dots which is W = 200 nm in both samples. Using the equation 3 with the experimentally obtained values of $\xi(0)$, we can find the temperature at which the wire network crossover should take place. Following this approach, we obtain that this crossover lies at 0.993$T_{c0}$ for SN sample and 0.994$T_{c0}$ in sample SCP, being $T_{c0}$ = 8.14 K and $T_{c0}$ = 8.45 K respectively.

Figure 2 shows the experimental phase diagram for the sample SN. Two different regimes are observed. At temperatures above the wire network crossover, the phase diagram follows the expected L-P parabolic background with $T_c$ (H) oscillations at specific H values (36). At temperatures below the wire network crossover the expected linear dependence is retrieved. Therefore, the L-P regime at temperatures close to $T_{c0}$ is confirmed, and the wire network crossover occurs at the expected temperature obtained by considering W = 200 nm. Interestingly, the SCP experimental data strongly deviates from the expected behavior considering that the L-P effects is only determined by the geometry of the nanostructured array. Figures 3(a) and 3(b) show the phase diagrams for the sample SCP in both, upward (a) and downward (b), magnetized states with the applied magnetic field parallel to the magnetization direction in each case. Here, the wire network crossover experimental value 0.988$T_{c0}$ is downshifted with respect to the expected value 0.994$T_{c0}$. This behavior resembles the magnetoresistance measured in SCP (see figure 1(b)), confirming that the stray field generated by the magnetic dots has a direct impact in the L-P regime. The magnetoresistance curves are shifted right or left depending on the direction of the applied fields and the out-of-plane remanent magnetization in the dots. This means that superconducting vortices and antivortices are present in this situation and they coexist with the L-P effect in the wire network regime (16-19). To figure out the exact origin of this magnetic influenced L-P effect we have to consider not only the vortex dynamics, but also the magnetism of the SCP dots and the direction of the applied fields, as it was discussed in the previous section. Gomez *et al.* (19) showed that effects related to vortex dynamics are present in the whole range of temperatures and they coexist with the L-P effect close to $T_{c0}$. The clear difference between the SN and SCP samples, although having the same geometry, indicates that the clue to understand the L-P effect in the SCP sample is the role played by the magnetic stray fields emerging from the out-of-plane dot magnetization.

In the usual case of a superconducting wire network with voids in presence of external magnetic fields, superconducting currents (supercurrents) circulate around the holes confining the magnetic flux inside them. In sample SCP, in addition to the external magnetic field, the stray field should be considered, since the stray field induces additional supercurrents even when no external magnetic field is applied. Figure 4 shows the simulated stray field generated by the Co/Pd dots in the SCP sample. The remanent configuration is obtained from the micromagnetic simulations.

As can be seen, the magnetic stray field expands beyond the dot dimension, giving place to a reduction of the superconducting channels (wires) to 145 nm ($W_{eff}$ in Fig. 4). When the applied magnetic field is parallel to the magnetization, the downshifted wire network crossover at $0.988T_{c0}$ gives a dot effective size of 256 nm. The effective broadening of the dots due to the stray-field explains perfectly the enhancement of the L-P temperature range. Therefore, the geometry of the wire network is not the single origin of the L-P effect; the L-P effect is governed and depends on the magnetic state of the dots. Remarkably, from the stray field simulations we can estimate that each magnetic dot creates one vortex-antivortex pair, which manifests in a phase diagram shifting of -129 Oe or +129 Oe, that corresponds to one fluxoid per dot, depending on the direction of the magnetization. For an external magnetic field antiparallel to the dots magnetization, that is, parallel to the stray field in the superconducting channels (wires), the asymmetry in the phase diagram results from a non-enhanced L-P regime.

Figure 5 shows that, for antiparallel external magnetic fields, the wire network crossover ($W_{(\uparrow\downarrow)}$) lies on the expected value, in contrast with the crossover ($W_{(\uparrow\uparrow)}$) found for parallel external magnetic fields. This result shows that, it is possible to manipulate the L-P effect with an external source, the direction of an applied magnetic field taking into account the direction of the out-of-plane remanent magnetization of the dots, upward or downward. Our results prove that magnetism can vary and control the crucial parameter W, which governs the wire network crossover and the rise of the L-P effect.

All these remarkable findings in SCP sample are in contrast with the situation in the SN sample. Despite the magnetic nature of the Ni dots, no asymmetry or shifting in the phase diagram is observed at all (Fig. 2). Moreover, the experimental value for the wire network crossover matches with the expected one $0.993T_{c0}$. Figure 6 depicts the stray field simulation for this sample. We observe that, in the case of Ni dots (magnetic vortex state)[28], the stray field remains within the dots, hence there is no enhancement of the effective size of the dot and no vortex-antivortex pairs emerge, giving rise to the usual geometrical L-P effect (see Fig. 2).

## Conclusions

We have studied the role played by magnetic stray fields generated by nanometric magnetic dot arrays in the development of Little-Park oscillations near the critical temperature. This has been achieved by comparing two arrays, with exactly the same unit cell, of magnetic nanodots embedded in a superconducting Nb film. In one array the dots exhibit a strong out of plane anisotropy, giving rise to a magnetic stray field that spreads out beyond the dot geometric boundary. In the other array (although having exactly the same dimensions) the magnetic dots are found in a vortex state, which results in a much weaker stray field localized on top of the dots.

Periodic oscillations and a quadratic dependence of critical temperature vs magnetic fields are observed for both samples in the (H,T) phase diagram, as clear evidence of a superconducting wire network regime. Usually, the wire network crossover only depends on geometric parameters, however a huge influence of the stray field has been found in the present case. The sample exhibiting out of plane anisotropy shows that an intense magnetic stray field enhances the L-P regime or not, depending on whether the applied magnetic field is parallel or antiparallel to the dot magnetization. The stray field profile and its alignment to the applied magnetic field have a direct impact in the wire network crossover parameters and therefore in the nature of the L-P regime. When the applied magnetic field is parallel to the dot magnetization, micromagnetic simulations permit us to elucidate that the crossover deviation is originated by an effective broadening of the size of the dots due to the magnetic stray fields. In this case, the stray field is high enough to create a vortex-antivortex pair, whereas when the magnetic dots are in the vortex state, the stray field is localized on top of the dots and does not generate any vortex-antivortex pair, giving rise to a thoroughly geometric L-P effect.

## Acknowledgments

We want to thank Spanish MICINN grants FIS2016-76058 (AEI/FEDER, UE), EU COST- CA16218. IMDEA Nanociencia acknowledges support from the 'Severo Ochoa' Programme for Centres of Excellence in R&D (MICINN, Grant SEV-2016-0686). MCO and AG acknowledges financial support from Spanish MICINN Grant ESP2017-86582-C4-1-R and IJCI-2017-33991; AMN acknowledges financial support from Spanish CAM Grant 2018-T1/IND-10360.

## Figure Captions

**Figure 1.**  Resistance (R) vs. applied magnetic field (H) measured for both samples at the same reduced temperature $T = 0.99T_{c0}$: (**a**) SN sample with the magnetic dots in vortex state; (**b**) SCP sample with the magnetic dots at remanence after applying a saturation magnetic field perpendicular to the sample. The red plot corresponds to a saturation field of $+1$ T while the blue plot corresponds to a $-1$ T saturation field.

**Figure 2.**  Reduced critical temperature $(T/T_{c0})$ vs. applied magnetic field (H) measured for SN sample (blue dots). Dotted horizontal line indicates the wire network crossover between the parabolic behavior (L – P regime) and the linear one. The red curve corresponds to the parabolic fit of the maxima observed in the experimental data (white dots).

**Figure 3.**  Reduced critical temperature $(T/T_{c0})$ vs. applied magnetic field (H) measured for SCP sample at remanence after applying a saturation magnetic field perpendicular to the sample: (**a**) negative magnetization ($-1$ T) and (**b**) positive magnetization ($+1$ T). Both figures show the region of the phase diagram that corresponds to the applied magnetic field parallel to the magnetization, when the wire network regime is enhanced by the stray field of the magnetic dots. Blue dots correspond to the experimental data, while the red curves are the parabolic fits to the maxima appearing in the experimental data (white dots).

**Figure 4.** Contour plot of the stray field calculated from the micromagnetic simulations at 50 nm on top of the Co/Pd dots for the magnetic SCP sample. Maximum and minimum values are noted next to the corresponding contour lines. Dotted black lines mark the geometric size of the dots, while $W_{eff}$ indicates the effective size of the superconducting wires obtained from the experimental phase diagrams. The color bar corresponds to stray-field values in Oe.

**Figure 5.** Phase diagrams measured for SCP sample at (**a**) negative magnetized and (**b**) positive magnetized states. Blue dots represent the experimental data. The red (green) curves correspond to the quadratic fit for the maxima (white dots) that appear at magnetic fields parallel (antiparallel) to the remanent magnetization of the dots. The superconducting wire network regime for the parallel case extends until the red dotted horizontal line ($W_{\uparrow\uparrow}$), is enhanced due to the stray field spreading beyond the dot. For the antiparallel case, the wire network regime crossover ($W_{\uparrow\downarrow}$) corresponds to an effective size that matches perfectly with the geometric profile of the dots.

**Figure 6.** Contour plot of the stray field calculated from the micromagnetic simulations at 50 nm on top of the Ni dots for the SN sample. Dotted black lines mark the geometric size of the dots. It is worth noting that the stray field generated by the dots in the SN sample is extremely confined, giving place to a situation in which the wire network crossover depends only on the geometric size of the nanostructures. The color bar corresponds to stray-field values in Oe.

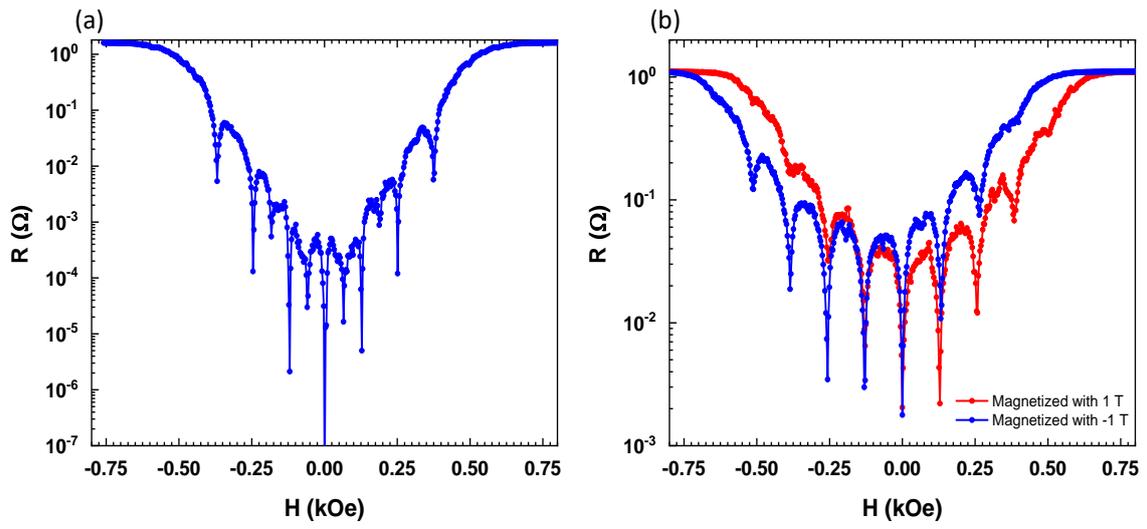

**FIGURE 1**

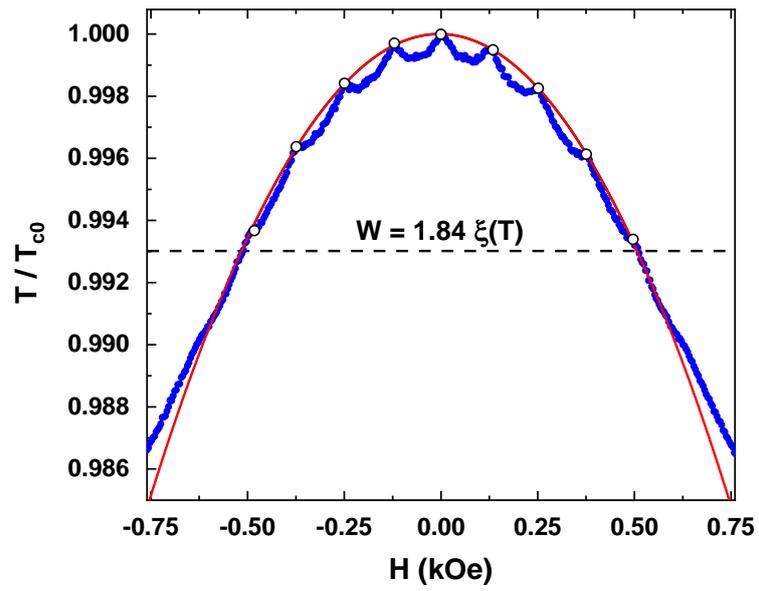

**FIGURE 2**

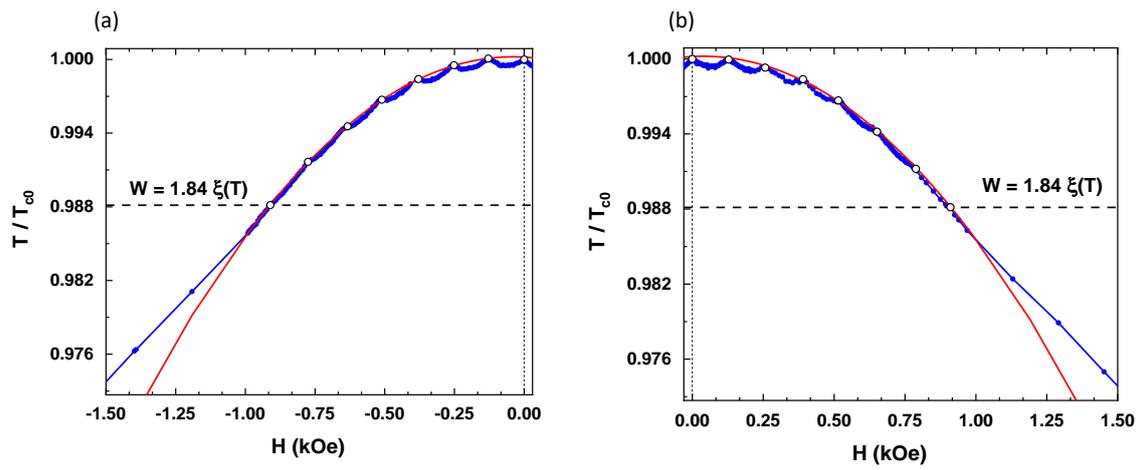

**FIGURE 3**

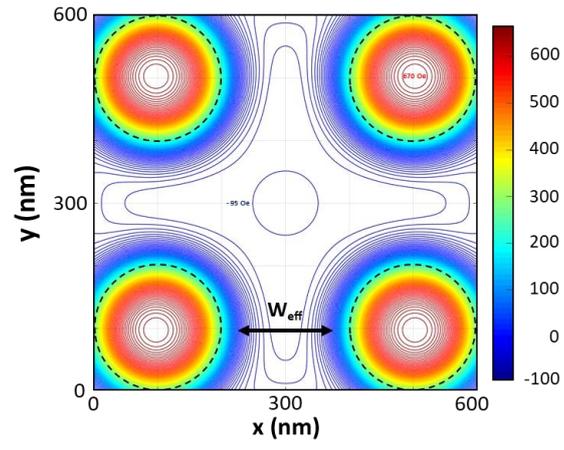

**FIGURE 4**

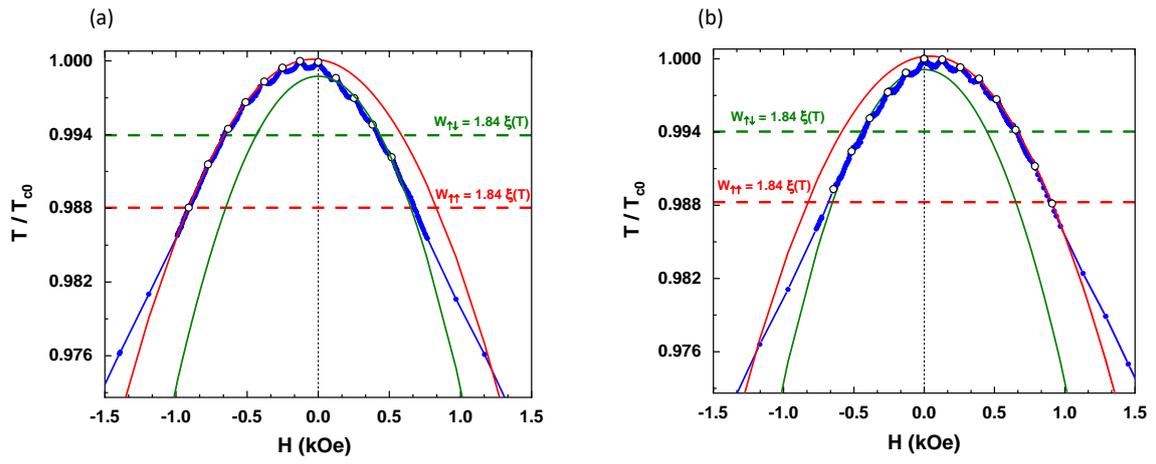

**FIGURE 5**

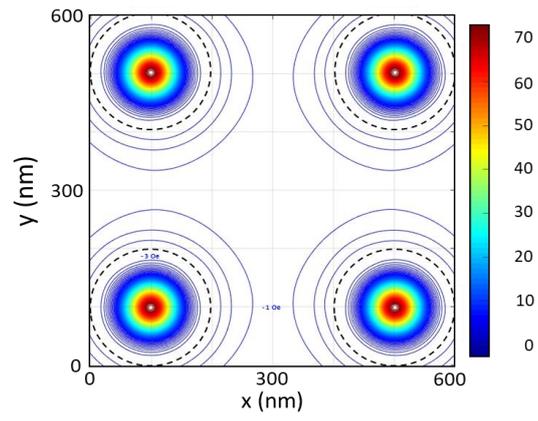

**FIGURE 6**